\documentclass{article}
\usepackage[latin1]{inputenc}
\usepackage[T1]{fontenc}
\usepackage[english]{babel}
\usepackage{amssymb}
\usepackage{amsmath}
\usepackage{amsthm}
\usepackage{graphicx}
\usepackage[margin=3cm]{geometry}
\usepackage{hyperref}
\hypersetup{
	colorlinks=true,
	linkcolor=blue,
	filecolor=magenta,      
	urlcolor=cyan,
}
\usepackage[font=small,labelfont=bf]{caption}
\usepackage{multirow}
\usepackage{authblk}
\usepackage{gensymb}
\usepackage{booktabs}
\usepackage{cleveref}
\usepackage{subcaption}

\DeclareMathOperator*{\argmin}{arg\,min}

\title{Crystalline phase discriminating neutron tomography using advanced reconstruction methods}

\author[1,2,*]{Evelina Ametova}
\author[3,4]{Genoveva Burca}
\author[5]{Suren Chilingaryan}
\author[6]{Gemma Fardell}
\author[4,7]{Jakob S. J\o{}rgensen}
\author[1,6]{Evangelos Papoutsellis}
\author[6]{Edoardo Pasca}
\author[1]{Ryan Warr}
\author[8]{\\Martin Turner}
\author[4]{William R. B. Lionheart}
\author[1]{Philip J. Withers}
\affil[1]{Henry Royce Institute, Department of Materials, The University of Manchester, M13 9PL, United Kingdom}
\affil[2]{Laboratory for Application of Synchrotron Radiation, Karlsruhe Institute of Technology, Germany}
\affil[3]{ISIS Pulsed Neutron and Muon Source, STFC, UKRI, Rutherford Appleton Laboratory, United Kingdom}
\affil[4]{Department of Mathematics, The University of Manchester, M13 9PL, United Kingdom}
\affil[5]{Institute for Data Processing and Electronics, Karlsruhe Institute of Technology, Germany}
\affil[6]{Scientific Computing Department, STFC, UKRI, Rutherford Appleton Laboratory, United Kingdom}
\affil[7]{Department of Applied Mathematics and Computer Science, Technical University of Denmark, Denmark}
\affil[8]{Research IT Services, The University of Manchester, M13 9PL, United Kingdom}

\affil[*]{Correspondence should be addressed to: evelina.ametova@kit.edu}

\begin{document}
\maketitle
\begin{abstract}

Time-of-flight neutron imaging offers complementary attenuation contrast to X-ray computed tomography (CT), coupled with the ability to extract additional information from the variation in attenuation as a function of neutron energy (time of flight) at every point (voxel) in the image. In particular Bragg edge positions provide crystallographic information and therefore enable the identification of crystalline phases directly. Here we demonstrate Bragg edge tomography with high spatial and spectral resolution. We propose a new iterative tomographic reconstruction method with a tailored regularisation term to achieve high quality reconstruction from low-count data, where conventional filtered back-projection (FBP) fails. The regularisation acts in a separated mode for spatial and spectral dimensions and favours characteristic piece-wise constant and piece-wise smooth behaviour in the respective dimensions. The proposed method is compared against FBP and a state-of-the-art regulariser for multi-channel tomography on a multi-material phantom. The proposed new regulariser which accommodates specific image properties outperforms both conventional and state-of-the-art methods and therefore facilitates Bragg edge fitting at the voxel level. The proposed method requires significantly shorter exposure to retrieve features of interest. This in turn facilitates more efficient usage of expensive neutron beamline time and enables the full utilisation of state-of-the-art high resolution detectors.

\end{abstract}
\section{Introduction} \label{sec:introduction}

Neutron radiography~\cite{maier1963use} and later neutron computed tomography (CT)~\cite{oien1977resonance,schlapper1977neutron} have been available at neutron sources for some time. As uncharged particles, neutrons probe the nucleus rather than the electron cloud and so in contrast to X-ray CT~\cite{maire2014quantitative} where the attenuation contrast increases with atomic number, neutron attenuation can give strong contrast between neighbouring elements in the periodic table (\emph{e.g.} Cu and Ni) or even different isotopes~\cite{strobl2009advances}. In addition to conventional attenuation contrast, the transmitted beam contains important information regarding coherent scattering from the crystalline lattice. Indeed, for many materials the scattering cross-section is comparable to the attenuation. Since the thermal neutrons produced by neutron spallation sources have wavelengths comparable to interplanar lattice spacings, polycrystalline materials will scatter the incident beam elastically according to Bragg's law. Since Bragg scattering can occur only for wavelengths shorter than twice the spacing between the lattice planes ($d_{hkl}$), the transmitted neutron spectrum exhibits characteristic abrupt increases in the transmitted intensity at these wavelengths. As a result the transmitted spectrum has a characteristic signature displaying distinct jumps in transmitted intensity corresponding to the Bragg edges for all the crystallographic phases in the material.

For pulsed spallation neutron sources it is possible to infer the energy and hence wavelength of each detected neutron from its time of flight (ToF) from the source to the detector. The development of pixelated ToF detectors enabled the Bragg edges to be imaged~\cite{santisteban2002strain,santisteban2002engineering}. The relation between spectral fingerprint and crystalline properties allows identification of polycrystalline materials and characterisation of their properties~\cite{santisteban2002engineering,kockelmann2007energy,kardjilov2011neutron} such as phase~\cite{steuwer2005using,song2017characterization}, texture and strain~\cite{santisteban2002strain,woracek2011neutron,wensrich2016bragg,reid2019application}. Combined with sample rotation, three-dimensional Bragg edge neutron CT is a natural extension of two-dimensional Bragg edge neutron imaging, as has been already demonstrated in several studies~\cite{woracek20143d,WATANABE2019162532,Carminati:ei5051}. However, as Kockleman et al.~\cite{kockelmann2007energy} pointed out in their early preliminary neutron energy selective imaging experiments, the signal, which is usually integrated over the white beam, becomes particularly low when distributed across many hundreds of ToF (energy) channels making tomographic imaging slow and noisy. In recent years the size of pixelated ToF discriminating detectors has improved significantly (from a $10 \times 10$ array of $2 \times 2~\mathrm{mm^2}$~\cite{santisteban2002strain} to a $512 \times 512$ array of $0.055 \times 0.055~\mathrm{mm^2}$~\cite{mcptremsin}), thereby improving the potential resolution, but further decreasing the flux received by each pixel and further confounding the reconstruction challenge. 

The conventional way of reconstructing tomographic datasets is filtered back-projection (FBP) which is a fast and well-established method, but demanding in terms of input data quality. As a result of the low count rates and hence slow acquisition, energy dispersive spectra tend to be heavily down-sampled prior to reconstruction and, even with reduced resolution, averaging over a relatively large region-of-interest in reconstructed images is still required to improve signal-to-noise ratio and characterise Bragg edges. Artificially induced low resolution thus currently hinders application of the technique. 

Here we present advanced reconstruction methods that make energy-dispersive neutron tomography a practical proposition. We explore the application of dedicated multi-channel reconstruction techniques with inter-channel correlation to improve reconstruction quality without compromising valuable resolution. Contrary to previous work~\cite{WATANABE2019162532,Carminati:ei5051} where each channel is treated as a separate reconstruction problem, we use iterative methods with prior information to jointly reconstruct spectral images. 

Iterative methods formulate reconstruction as an optimisation problem consisting of a data fidelity term and a regulariser. The latter term encodes prior knowledge about the image and helps to find a unique solution for the ill-posed tomographic problem by encouraging desired properties in reconstructed images. There is no unique regulariser which performs well for all tomographic problems and it has to be carefully selected depending on the underlying tomographic data properties. In this study we tailored a regulariser specifically for Bragg edge neutron CT based on a combination of Total Variation (TV) regularisation~\cite{rudin1992nonlinear,sidky2006accurate} in the spatial dimension and Total Generalised Variation (TGV) regularisation~\cite{bredies2010total,kazantsev2016sparsity} in the spectral dimension. TV preserves edges and suppresses noise by encouraging piece-wise constant regions in reconstructed spatial images~\cite{rudin1992nonlinear,sidky2006accurate}, while TGV regularisation promotes characteristic piece-wise smooth behaviour in the spectral dimension~\cite{bredies2010total}. For brevity, henceforth we refer to the proposed method as TV-TGV. 

We compare TV-TGV against conventional FBP reconstruction and a state-of-the-art regulariser for multi-channel CT images -- Total Nuclear Variation (TNV). TNV is a recent regulariser which enforces common edges across all channels in multi-channel images~\cite{holt2014total,rigie2015joint,kazantsev2018joint}. Wider implementation of these advanced reconstruction methods is possible through the open source CCPi Core Imaging Library (CIL) reconstruction framework~\cite{jorgensen2021cil,papoutsellis2021cil}, for example one may employ TV only or TGV only if desired. The performance of all methods is demonstrated on a multi-material sample comprising aluminium cylinders filled with various metallic powders of high purity.

Experimental details with a particular focus on preprocessing are presented in~\cref{sec:experiment}. In~\cref{sec:methods}, we provide a general overview of reconstruction methods used in the present study. In~\cref{sec:results}, a comparison between all the reconstruction methods in this study is presented. A Bragg edge fitting procedure is further used to detect and locate Bragg edges in TNV and TV-TGV reconstructions and extract crystallographic information. We also demonstrate decomposition of reconstructed spectral images into individual material maps. Discussion and conclusions are given in~\cref{sec:conclusion}.

\section{Experimental methods} \label{sec:experiment}

\subsection{Measurement model}

The measurement model in ToF neutron CT is given by the Beer-Lambert law, which relates attenuation properties of material to measured intensity:

\begin{equation} \label{eq:beer_lambert}
	I = I_0 \exp \left( -\int_L \mu(x, \lambda) \mathrm{d}x \right),
\end{equation}

\noindent where $I_0$ and $I$ corresponds to the beam intensity incident on the object and on the detector element, respectively, $L$ is a linear path through the object and $\mu(x, \lambda)$ is the wavelength-dependent attenuation coefficient at the physical position $x$ in the object for the given wavelength $\lambda$. The probability of neutron-matter interaction is a function of neutron energy (wavelength) and is given by the microscopic total cross-section $\sigma_{tot} (\lambda),~[\mathrm{cm}^2]$ of the nucleus. When neutrons travel through material, the probability of interaction depends not only on the microscopic total cross-section $\sigma_{tot} (\lambda)$, but also on the number $N,~[\mathrm{atoms/cm}^3]$ of nuclei within unit volume of material. The macroscopic total cross-section $\Sigma_{tot} (\lambda),~[\mathrm{cm}^{-1}]$ defines the probability of neutron-matter interaction per unit distance travelled of the neutron~\cite{fundamentals1993nuclear}, \emph{i.e.} the attenuation coefficient:

\begin{equation}
	\mu(\lambda) = \Sigma_{tot} (\lambda) = N \sigma_{tot} (\lambda)
\end{equation}

\begin{figure}[h!]
	\centering
	\includegraphics[width=5in,height=5in,keepaspectratio]{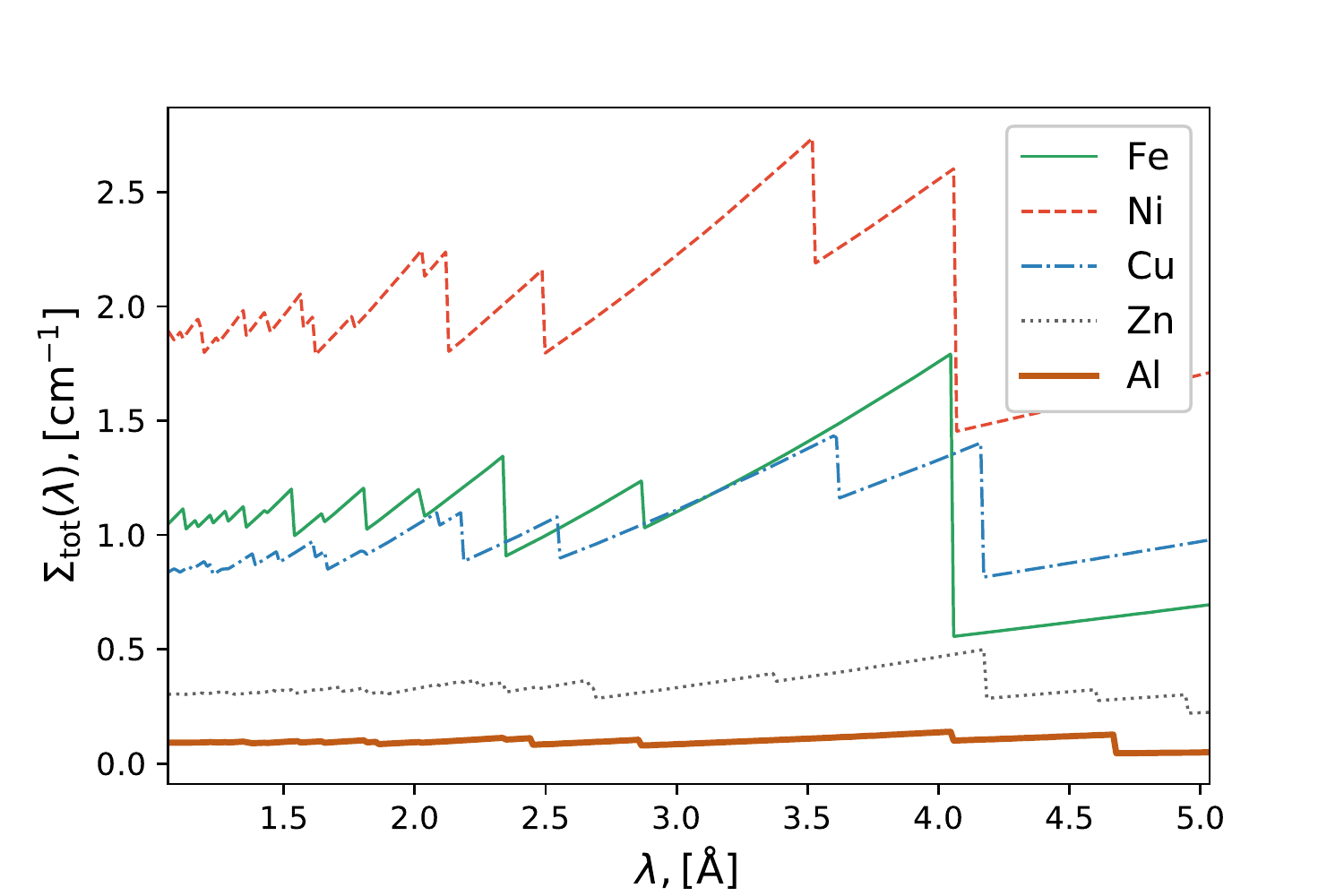}
	\caption{Theoretical neutron spectra for materials employed in this study (Calculated using the NXS software package~\cite{Boin:db5100}).}
	\label{fig:neutron_transmission}
\end{figure}

The microscopic total neutron cross-section $\sigma_{tot} (\lambda)$ is a linear combination of several contributions, defining the probability of elastic, inelastic, coherent and incoherent scattering, and absorption~\cite{boin2011validation}. All interaction processes contribute to the decrease of transmitted intensity, however only coherent elastic scattering is responsible for characteristic abrupt increases in the transmitted intensity at $2d_{hkl}$ allowing the detection of specific lattice planes. In \cref{fig:neutron_transmission} we show the wavelength-dependent macroscopic cross-section $\Sigma_{tot} (\lambda)$ for materials employed in this study. 

\subsection{Sample design} \label{subsec:sample}
\begin{figure}
	\centering
	\includegraphics[width=4in,height=4in,keepaspectratio]{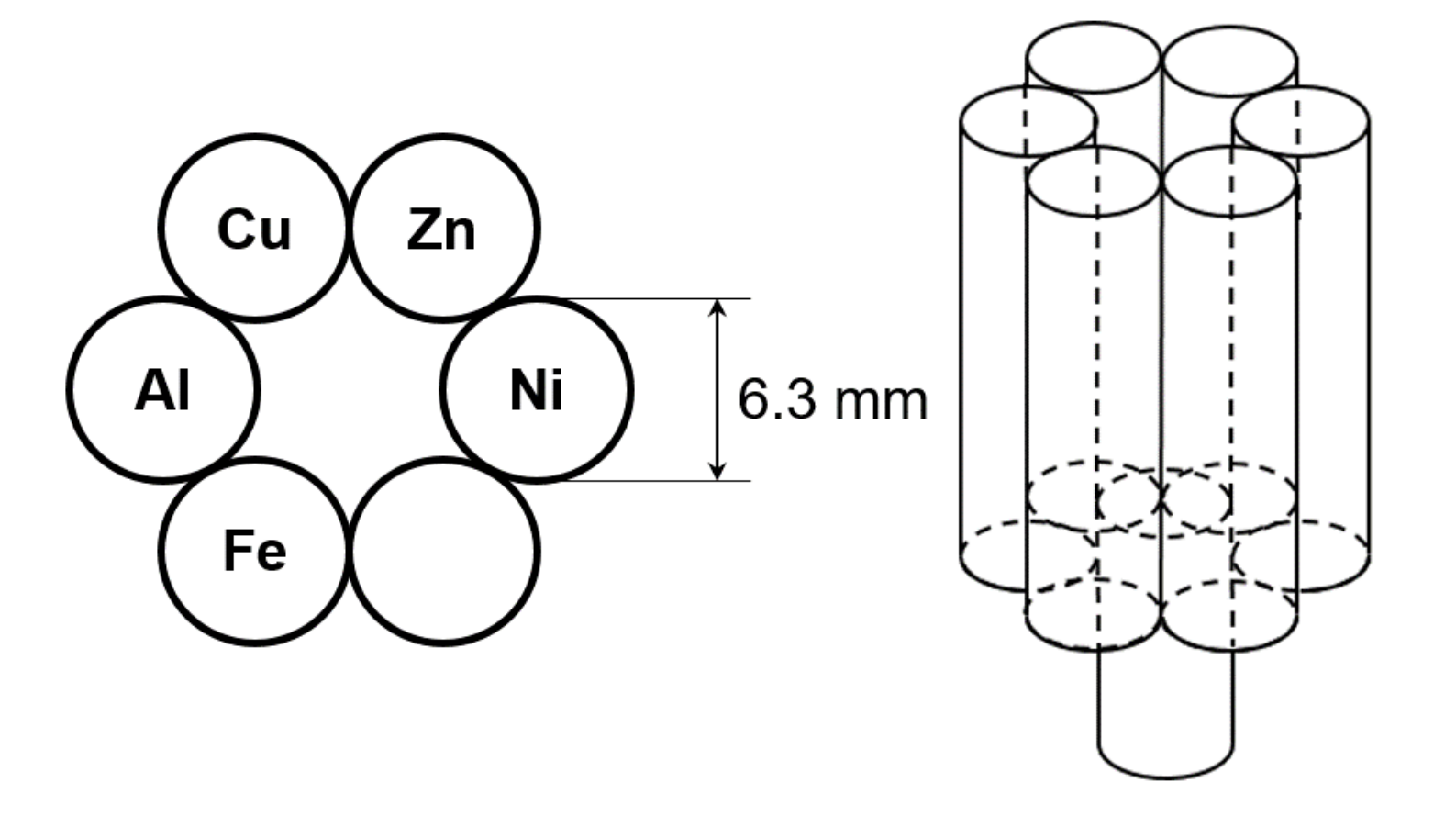}
	\caption{Sketch showing the test sample employed in the study comprising Al, Fe, Cu, Ni and Zn powders.}
	\label{fig:sample}
\end{figure}

For this study, a sample was constructed comprising six 6.3~mm diameter thin-walled cylindrical containers formed from aluminium foil. Five were filled with metal powders, namely copper (Cu), aluminium (Al), zinc (Zn), iron (Fe) and nickel (Ni); one cylinder was left empty. The containers were sealed and affixed around a hollow aluminium cylinder in a hexagonal close packed arrangement (\cref{fig:sample}). Powders were chosen for the sample to reduce the effect of texture which can strongly affect the shape of the Bragg edges~\cite{xie2018applying}. This provides idealised Bragg edge spectra likely to be in good agreement with theoretical calculations of neutron transmission (\cref{fig:neutron_transmission}). 

\subsection{Instrument settings} \label{subsec:settings}

The data was acquired at the Imaging and Materials Science \& Engineering (IMAT) beamline operating at the ISIS spallation neutron source (Rutherford Appleton Laboratory, U.K.)~\cite{jimaging4030047,Burca_2013}. IMAT operates in a ToF measurement mode. Neutrons generated through spallation are slowed down by the L-H2 moderator on IMAT, so they become ``thermalised''. Neutrons reach the sample and then the detector at different times according to their energy. Neutrons with shorter wavelength have higher velocity and are recorded first followed by increasingly less energetic neutrons. Consequently, given the distance travelled between the source and the detector, and the elapsed time from the pulse leaving the source the energy, and hence wavelength, can be calculated based on the de Broglie equation.

IMAT is installed on ISIS TS2 (2nd target station) which operates at 10~Hz repetition rate and the flight path between the source (more specifically, neutron moderator) and the detector is $\approx56.4$~m giving an effective wavelength range around 6~{\AA}. The IMAT beamline is equipped with a borated-microchannel plate (MCP) detector combined with Timepix chip~\cite{mcptremsin}. The detector consists of a $2 \times 2$ array of $256 \times 256$ readout chips, resulting in $512 \times 512$ active pixels. The pixel size is 0.055~mm giving a field of view of approximately $28 \times 28~\mathrm{mm}^2$. The ToF spectrum recorded at each pixel can have up to 3100 time (energy) bins. There are small gaps and a slight misalignment between the readout chips~\cite{mcptremsin}, but these are not expected to have any significant implications for the current study. 

The MCP detector functions in event timing mode, in which the arrival time of each neutron at each pixel is measured with respect to some external trigger (pulse). Unfortunately, the detector can register only one event per pixel per time frame. Consequently, slower (lower energy) neutrons have a lower probability of being detected, as some pixels are already occupied by faster neutrons. This effect is typically referred to as detector dead time. It is possible to set-up several time frames (also referred to as shutter intervals) within one pulse with arbitrary length and bin width for each time frame, with data read out in the end of each shutter interval. Data readout takes 320~$\mu$s and introduces gaps in the measured ToF signal. The introduction of several shutter intervals within one pulse reduces the spectral signal distortions due to detector dead time, but cannot fully eliminate it. In~\cite{Tremsin_2014} authors proposed an algorithm called ``overlap correction'' to compensate for the counts lost. The efficiency of the algorithm was confirmed in~\cite{WATANABE201755}. For the present study, the detector shutter intervals were selected in such a way that data readout take place between theoretical Bragg edges for all measured materials (\cref{table:shutter_intervals}). With this configuration 2843 energy channels were measured for every projection having wavelength resolutions between $0.7184 \cdot10^{-3}$~{\AA} and $2.8737 \cdot10^{-3}$~{\AA}.

\begin{table}[h]
	\centering
	\begin{tabular}{ c c | c c | c c | c}
		\toprule
		\multicolumn{2}{ c| }{Beginning} & \multicolumn{2}{ |c| }{End} & \multicolumn{2}{ |c| }{Bin width} & \multirow{2}{*}{Number of channels} \\
		\cline{1-6}
		\\[-1em]
		ms & {\AA} & ms & {\AA} & $\mu$s & {\AA} \\
		\midrule
		15.0000 & 1.0524 & 26.6800 & 1.8719 & 10.2400 & $0.7184 \cdot10^{-3}$ & 1141 \\
		27.0000 & 1.8943 & 43.6800 & 3.0646 & 20.4800 & $1.4369 \cdot10^{-3}$ & 814 \\
		44.0000 & 3.0870 & 52.6800 & 3.6960 & 20.4800 & $1.4369 \cdot10^{-3}$ & 424 \\
		53.0000 & 3.7185 & 72.0000 & 5.0515 & 40.9600 & $2.8737 \cdot10^{-3}$ & 464 \\
		\bottomrule
	\end{tabular} 
	\caption{MCP detector shutter intervals chosen for the experimental study.}\label{table:shutter_intervals}
\end{table}

\subsection{Data acquisition}

The sample (\cref{fig:sample}) was placed and clamped onto the rotary stage. The sample was moved as close as possible to the MCP detector using the kinematic system. An alignment laser was used to visually align the vertical axis of the sample with respect to the vertical edge of the detector and to ensure that the object was within the field of view for all rotation angles. A set of spectral projections were acquired at 120 equally-spaced angular positions over $180\degree$ rotation ($1.5\degree$ angular increments). Each projection was acquired with 15~min exposure time.

We acquired two sets of flat field (also referred to as open beam) images (before and after acquisition) to compensate for detector imperfections and decrease of beam intensity over time which was observed in other experiments. Also, flat field averaging is generally recommended to reduce ring artifacts. Therefore 4 spectral flat field images were acquired before and 4 after sample acquisition (8 in total), each one with 15 min exposure. The MCP detector has no dark current noise. The raw data is available at~\cite{jorgensen2019tof}.

\subsection{Data preprocessing} \label{subsec:data_preprocessing}

Following~\cite{Tremsin_2014,Liptak_2019}, MCP detector related corrections (including both overlap correction and scaling to the same number of incident neutrons) were performed both for the set of 120 projections and the set of 8 flat field images. We performed three different types of flat-field correction: 

\begin{itemize}

\item Using an average of 4 flat field images acquired before sample data acquisition.

\item Using 2 averaged flat field images, where the first flat field image corresponds to an average of 4 flat field images acquired before the data acquisition and the second one corresponds to an average of 4 flat field images acquired afterwards. Then, the actual flat field image for every projection is calculated based on pixel-wise channel-wise linear interpolation of intensity values between 2 averaged flat field images.

\item Using an average of all 8 flat field images. The intensity of the averaged flat field image is then multiplied by the quotient of the mean intensity in unoccupied detector columns in a projection image and mean intensity in the averaged flat field image itself. The scaling factor is calculated individually for each projection and each channel. We refer to this approach as \emph{flux normalisation}. 
	
\end{itemize}

\begin{figure}
	\centering
	\begin{subfigure}[b]{0.45\textwidth}
		\centering
		\includegraphics[clip,trim=1.75in 0in 2.75in 0in,width=3in,height=3in,keepaspectratio]{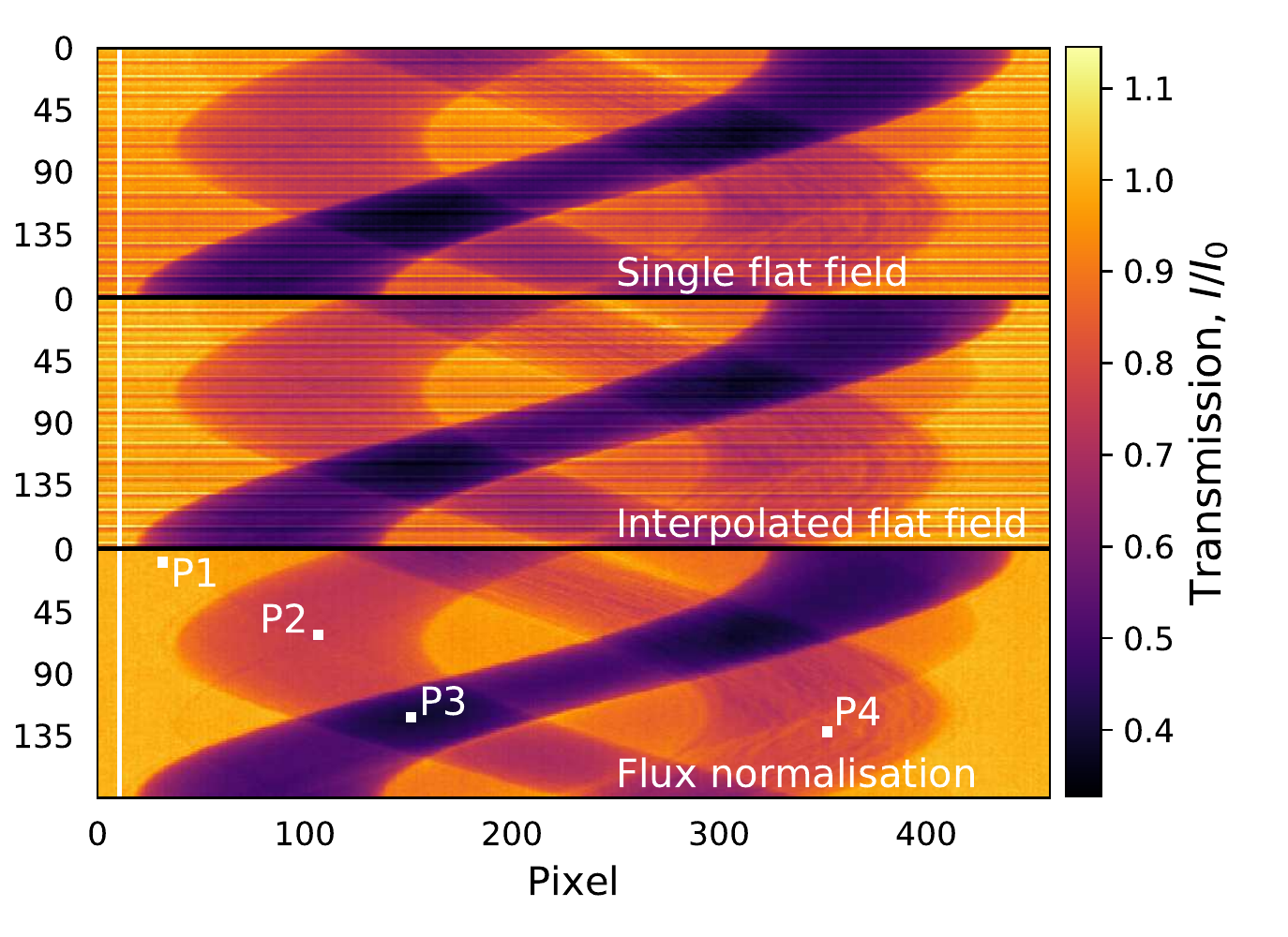}
		\caption{ }
		\label{subfig:flat_field_comp}
	\end{subfigure}
	\hfill
	\begin{subfigure}[b]{0.45\textwidth}
		\centering
		\includegraphics[width=3in,height=3in,keepaspectratio]{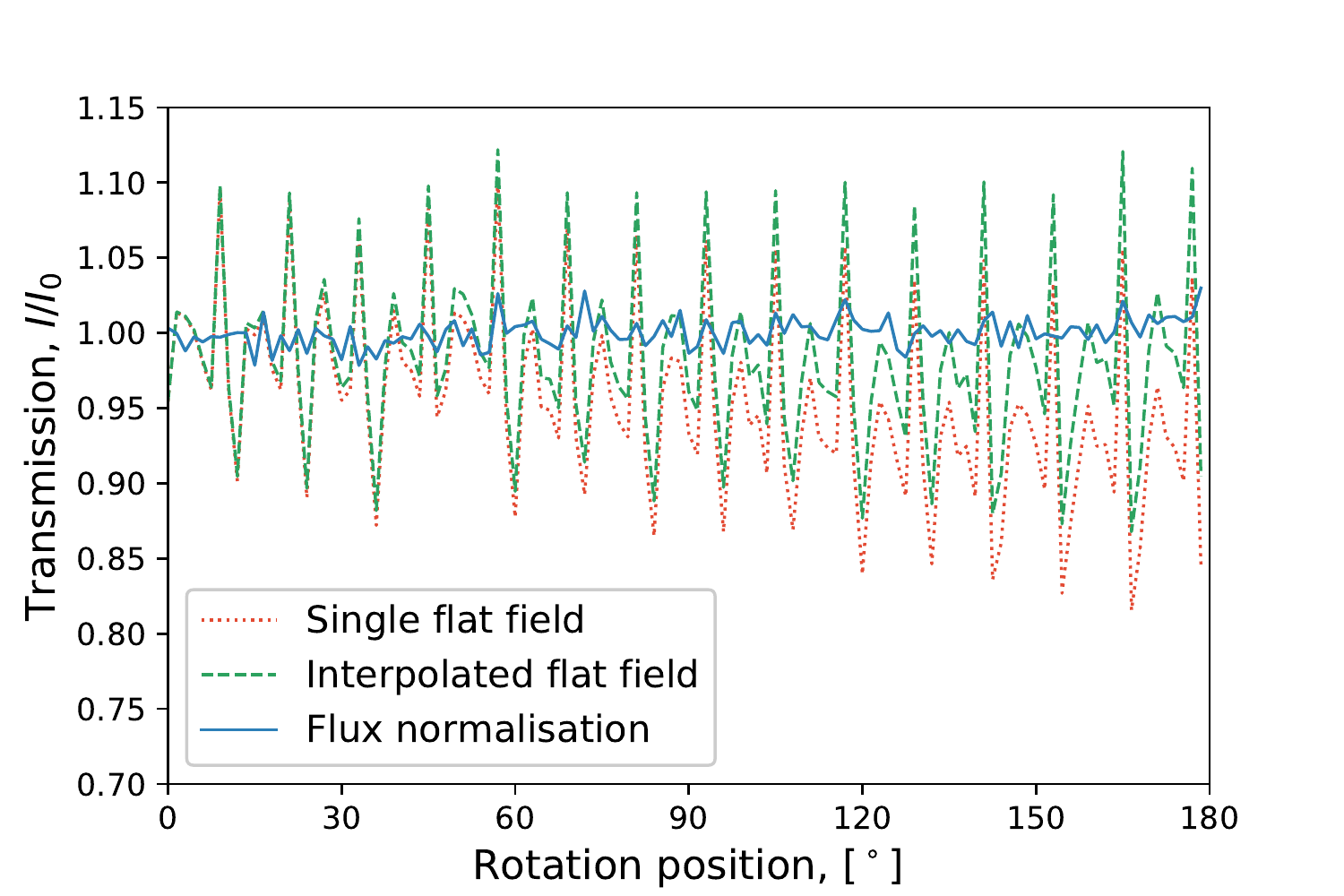}
		\caption{ }
		\label{subfig:flat_field_correction}
	\end{subfigure}
	
	\caption{Effect of flat field correction. a) Flat-field corrected white beam (sum of all wavelength) transmission sinogram for a single slice of the acquired dataset. The white vertical line indicates the location of the profile line. b) Intensity profile along vertical profile lines. P1-P4 label pixels chosen for examination of acquired spectra in the next figure.}
	\label{fig:flat_field}
\end{figure}

To demonstrate the effect of three different flat field correction approaches we show a flat-field corrected white beam (sum of all wavelength) transmission sinogram for a single slice of the acquired dataset (\cref{fig:flat_field}). Noticeable horizontal stripes are present for the first two flat field correction approaches which are eliminated with the flux normalisation. Profile lines (marked as a white solid line in the surrounding air in~\cref{subfig:flat_field_comp}) spotlight the difference between the first two approaches: linear interpolation compensates for the upwards trend in image intensity (\cref{subfig:flat_field_correction}). Finally, the flux normalisation not only de-trends the signal but also drastically reduces the overall fluctuations. As a result, the flux normalisation approach was used in the present study.

As relatively short exposure time was used for the acquisition, spectral images were down-sampled in the spectral dimension to increase counting statistics and produce reasonable spectra. Down-sampling was performed for each shutter interval independently by taking the average of each 16, 8, 8 and 4 channels for the respective shutter intervals (\cref{table:shutter_intervals}). As a result, each preprocessed projection contains 339 channels having a uniform bin width of $11.5 \cdot10^{-3}$~{\AA} over a wavelength range between 1~{\AA} and 5~{\AA}.
 
\begin{figure}
	\centering
	\begin{subfigure}[b]{0.45\textwidth}
		\centering
		\includegraphics[clip,width=3in,height=3in,keepaspectratio]{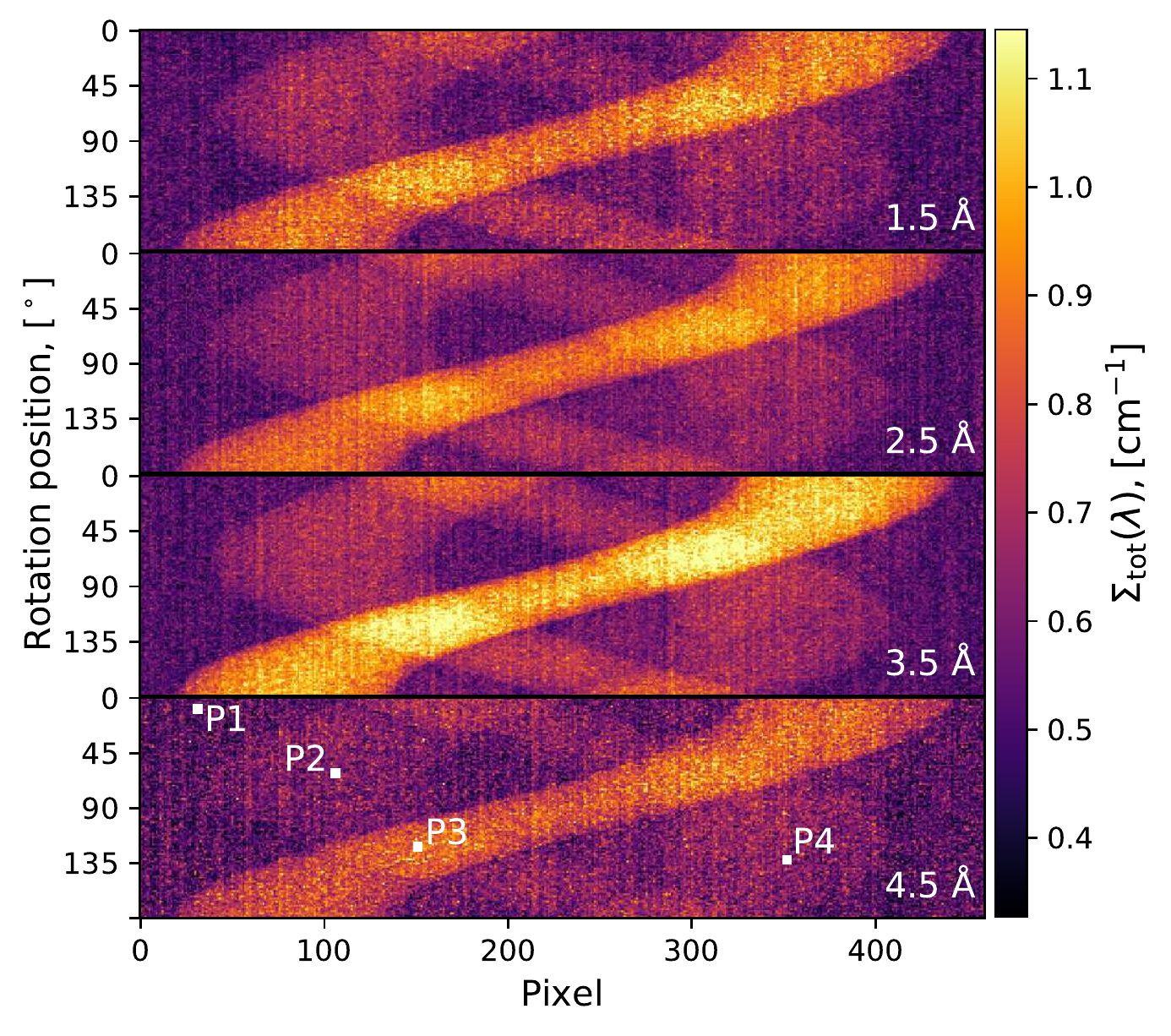}
		\caption{ }
		\label{subfig:sino_spectral}
	\end{subfigure}
	\hfill
	\begin{subfigure}[b]{0.45\textwidth}
		\centering
		\includegraphics[width=3in,height=3in,keepaspectratio]{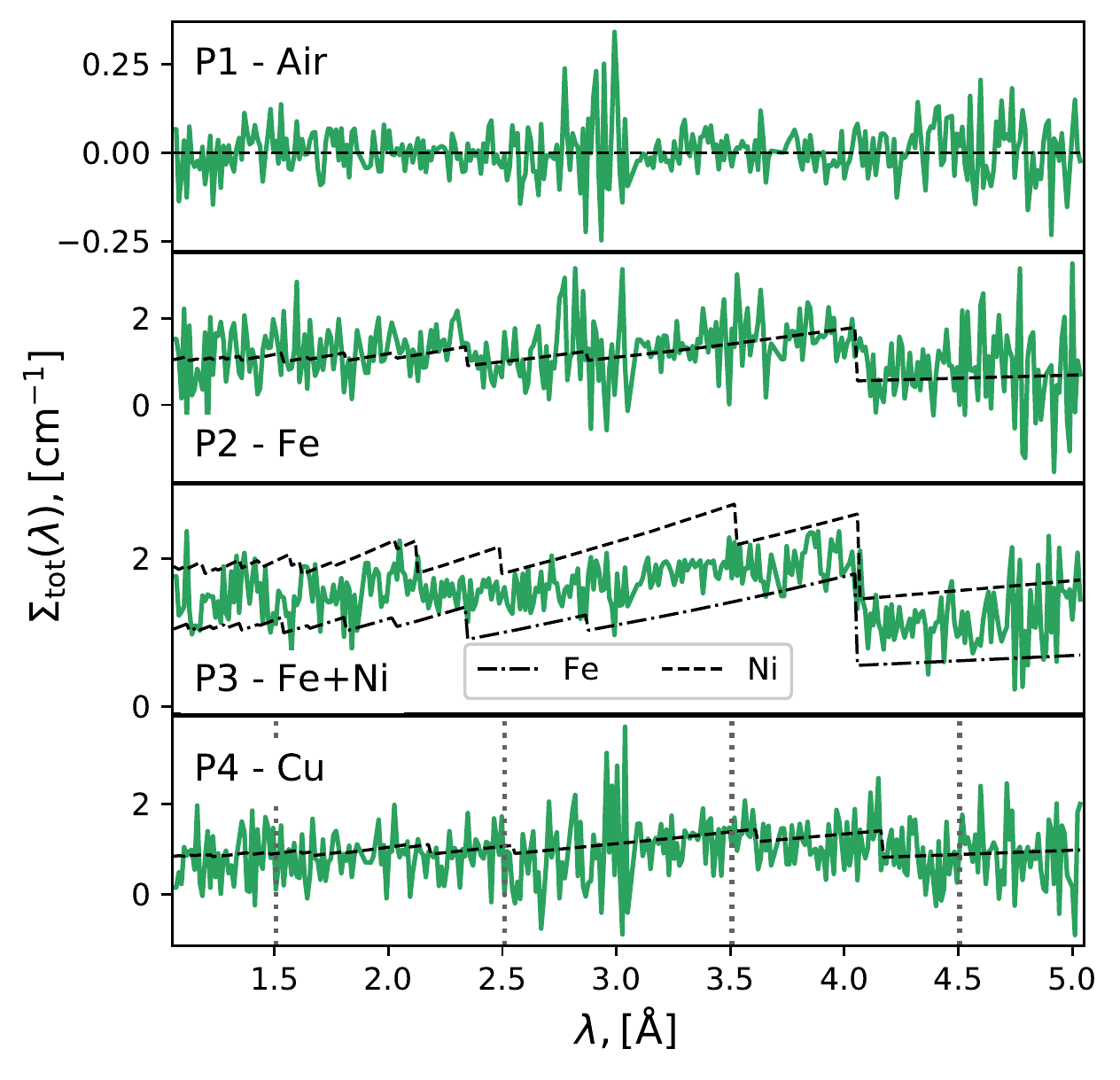}
		\caption{ }
		\label{subfig:signal_spectral}
	\end{subfigure}
	
	\caption{Sinograms (a) and recorded spectra (b). The spectra have been normalised by the transmission path length through the corresponding material. P1-P4 in (a) denote the location of pixels chosen for examination of recorded spectra. Vertical dotted lines in the bottom subplot in (b) mark the channels chosen for visualisation in (a).}
	\label{fig:sino}
\end{figure}

In~\cref{subfig:sino_spectral}, we show the preprocessed attenuation sinograms for individual wavelength channels. Even after spectral downsampling, strong noise is present in the measured tomographic data. Recorded spectra (\cref{subfig:signal_spectral}) taken for a path through surrounding air (P1), iron (P2), both iron and nickel (P3) and copper (P4) show that noise dominates over valuable Bragg edge information and makes Bragg edge characterisation for individual pixels barely feasible. 

It is also important to recognise that the incident spectrum on IMAT is a function of the neutron moderation process. The incident spectrum has a crude ``bell shape'' with a peak around 2.6~{\AA}~\cite{jimaging4030047,Burca_2013}. The recorded spectrum is also affected by the detector dead time. As a result the number of counts, and consequently noise level, in each channel depends on wavelength and position of shutter intervals. The elevated noise level is noticeable in the wavelength channels below 1.5~{\AA} and above 4.5~{\AA}, where IMAT has the lowest flux, and between 2.5~{\AA} and 3~{\AA} where a combination of high flux and detector dead time causes distortions in the recorded spectrum. The wavelength-noise dependence has direct implications on the reconstruction quality of each individual channel. 

Short of further downsampling of the recorded multi-channel images, advanced reconstruction methods are needed to handle noisy multi-channel neutron CT data. A potential solution to this low-count CT problem is to effectively exploit available prior information, such as the anticipated image properties and structural correlation between channels. 

\section{Methods} \label{sec:methods}

\subsection{Tomographic reconstruction}

Tomographic reconstruction aims to recover a map of sample attributes from a set of integral measurements acquired at various angles. In every wavelength channel, Bragg edge neutron CT can be well approximated by the standard absorption tomography model (Radon transform). Consequently, algorithms developed for X-ray CT can be conveniently used to reconstruct the neutron attenuation map for every wavelength channel. Here, we focus on the two-dimensional reconstruction problem in the spatial domain; extension to the third spatial dimension is straightforward. 

Given a multi-channel tomographic dataset with $K$ channels, a conventional approach is to reconstruct each channel independently using the FBP algorithm. The FBP algorithm is derived from the Fourier Slice theorem which relates line integral measurements to two-dimensional Fourier transform of an object's slice. In FBP-type reconstruction methods, projections are filtered independently and then back-projected onto the plane of the tomographic slice. FBP reconstruction is very fast but requires high quality input data and dense angular sampling to achieve good results.

Alternatively the inverse problems framework can be used to reconstruct low-count CT data. Consider a single channel $k, k = 0,1 \dots, K-1$, in a spectral dataset consisting of $K$ wavelength channels, and let $\overline{b}(\lambda_k)$ be a recorded discrete sinogram with $P \times D$ elements, with $P$ being the total number of projections and $D$ being the number of pixels in a detector row. The sinogram $\overline{b}(\lambda_k)$ is vectorised as a column vector with $M = P D$ elements (obtained by stacking the columns in the original two-dimensional sinogram). Let $\overline{u}(\lambda_k)$ be a vectorised two-dimensional array of material attributes we want to reconstruct with $N = D^2$ elements (voxels). The discrete version of the Radon transform is then given by the projection operator $\overline{A}$ containing $M \times N$ elements. If $i, i = 0,1, \dots M-1$ and $j, j = 0,1, \dots ,N-1$, then $\overline{A}_{(i,j)}$ is the length of intersection of the $i$.th ray with the $j$.th voxel. The reconstruction problem then takes a form:

\begin{equation} \label{eq:au_eq_b}
\overline{b}(\lambda_k) = -\textrm{ln} \left( \frac{I(\lambda_k)}{I_0(\lambda_k)} \right) \approx \overline{A} \overline{u}(\lambda_k).
\end{equation}

\noindent where $I_0(\lambda_k)$ and $I(\lambda_k)$ corresponds to the wavelength-dependent flux measured without (open beam) and with a sample in the field of view, respectively. 

Data acquisition in Bragg edge neutron CT is very time consuming since neutron fluxes are typically low (compared to synchrotron X-ray sources) and because the detected neutrons are shared among multiple time of flight (energy) channels. Hence, the acquisition mode does not inherently provide sufficient data for a numerically stable solution of~\cref{eq:au_eq_b}, similar to the low-dose tomography problem in medical X-ray CT imaging. The resulting problem is said to be ill-posed in a mathematical sense, \emph{i.e.} the naive solution of~\cref{eq:au_eq_b} will barely produce any useful result. A common treatment for ill-posed problems is to design a surrogate problem which is consistent with the original problem but is well-posed and computationally tractable. In other words, we obtain a stable approximate solution by incorporating some prior knowledge about the problem. 

Multi-channel CT data can be considered as a stack of two- or three-dimensional datasets, where every voxel in the reconstructed volume contains a vector with a spectral material response, spanning an additional dimension. In this respect, every data point in the material response vector belongs in the same physical structure, \emph{i.e.} channels share structural information. Then, inter-channel correlations can be utilised to improve reconstruction quality. The spectral CT reconstruction takes a form:

\begin{equation}
	b = A u,
\end{equation}

\noindent where $b$ and $u$ are obtained by stacking $K$ column vectors $\overline{b}(\lambda_k)$ and $\overline{u}(\lambda_k)$, respectively, and $A = I_{(K \times K)} \otimes \overline{A}$, $\otimes$ is the Kronecker product, and $I_{(K \times K)}$ is the identity matrix of order $K$. 

The reconstruction problem is then constructed as an optimisation problem:

\begin{equation} \label{eq:optimisation_problem}
\argmin_u \{ F(u) = f(b,Au) + \alpha g(u) \},
\end{equation}

\noindent where $f(b,Au)$ is a data fidelity metric which measures the discrepancy between the projection of solution $u$ and the acquired data $b$. The regularisation term $g(u)$ imposes certain prior assumptions typically expressed in terms of desired image properties. The scalar parameter $\alpha$ controls the trade-off between fit with the acquired data $u$ and the regularisation. The choice of regulariser depends on the underlying problem since there is no unique regulariser which performs best in all problems. The art of choosing the right regulariser and finding a balance between the two terms is a challenging task, as regularisation inevitably introduces bias into the solution. The bias is a price one pays for solving a problem which is not solvable by other means~\cite{jorgensen2013sparse}.

TV is the most commonly used regulariser. TV encourages a sparse image gradient and consequently favours piece-wise constant images with sharp boundaries~\cite{rudin1992nonlinear,sidky2006accurate}. The TV model describes well the spatial dimension of images of the type to be reconstructed in this study, but it is known to introduce ``staircase'' artifacts for piece-wise affine signals, \emph{i.e.} a ramp or features similar to the ones observed in wavelength dependent attenuation coefficient (\cref{fig:neutron_transmission}) might be reconstructed as piece-wise constant, reminiscent of a staircase. TV is also known to suffer from intensity reductions and, as a result, low contrast regions might be lost. Multiple modifications of TV have been proposed to cure the above shortcomings.

The TNV approach proposed in~\cite{rigie2015joint} explicitly promotes reconstructions with common edges across all channels. The idea is based on the fact that having shared gradient directions is equivalent to have a rank-one Jacobian of a multi-channel image. Therefore, TNV penalises the singular values of the Jacobian. In the spatial dimension, TNV has similar properties to TV regularisation, \emph{i.e.} it also favours a sparse image gradient. Consequently, TNV correlates channels and improves reconstruction quality by promoting common structures in multichannel images. Application of TNV for reconstruction of multi-channel images has been successfully demonstrated in~\cite{rigie2015joint,kazantsev2018joint,zhong2018eds}. The reconstruction problem is then formulated as,

\begin{equation} \label{eq:tnv_recon}
F(u) = \| Au-b \|^2_2 + \alpha \mathrm{TNV}(u).
\end{equation}

Similar to TV, TNV suffers from a loss of contrast. Secondly, TNV does not allow the decoupling of regularisation parameters for the spatial and spectral dimensions which makes it impossible to balance the level of regularisation between dimensions.

Here we propose a novel tailored regulariser which treats the spatial and spectral dimensions separately. As the TV model captures piece-wise image properties in the spatial dimension, we rely on another regulariser to support reconstruction in the spectral dimension. A natural remedy for the staircase artifact is to use higher order derivatives in the regularisation term. Here, we use TGV~\cite{bredies2010total} which allows for both sharp changes in spectral signal and gradual intensity changes. TGV also effectively exploits inter-channel correlation. In this case the reconstruction problem is formulated as,

\begin{equation} \label{eq:tv_tgv_recon}
F(u) = \| Au-b \|^2_2 + \beta \mathrm{TV}_{x,y} (u) + \gamma \mathrm{TGV}_c (u).
\end{equation}

\noindent Here we use $\mathrm{TV}_{x,y}$ to designate a TV operator over two spatial dimensions $x$ and $y$, whereas $\mathrm{TGV}_c$ operates over channel dimension. $\mathrm{TV}_{x,y}$ is a channel-by-channel operator applied to all channels individually and then summed over all channels. $\mathrm{TGV}_c$ is a one-dimensional operator applied in each individual voxel across all channels simultaneously. 

TGV regularisation has been already successfully applied to multi-channel tomographic data~\cite{knoll2016joint,schloegl2017infimal,holler2018coupled,huber2019total,chatnuntawech2016vectorial} including time resolved magnetic resonance imaging (MRI), energy-dispersive X-ray spectroscopy (EDXS) and high-angle annular dark field (HAADF) data. To the best of our knowledge, this is the first attempt to combine TV and TGV for spectral tomographic problems, in this case Bragg edge neutron CT data. The decoupling of regularisation into the spatial and spectral dimension allows us to balance the two regularisation terms appropriately and promote the desired properties in both dimensions independently. However, this greater flexibility comes at higher computational cost. The more terms that are included into the optimisation problem~\cref{eq:optimisation_problem}, the less it is suitable for parallelisation and software acceleration.


\subsection{Numerical implementation}

A common way to solve the large-scale optimisation problem in~\cref{eq:optimisation_problem} is by means of the Fast Iterative Shrinkage-Thresholding Algorithm (FISTA)~\cite{beck2009fast} or the Primal-Dual Hybrid Gradient (PDHG) algorithm~\cite{chambolle2011first}. Here we used the implementation of both FISTA and PDHG  in the CCPi Core Imaging Library (CIL)~\cite{jorgensen2021cil,papoutsellis2021cil}. CIL provides a highly modular Python library for prototyping of reconstruction methods for multi-channel data such as spectral and dynamic (time-resolved) CT. CIL wraps a number of third-party libraries with hardware-accelerated building blocks for advanced tomographic algorithms. CIL relies on the ASTRA toolbox~\cite{van2016fast,VANAARLE201535,PALENSTIJN2011250} to perform forward- and back-projection operations and provides a set of various regularisers through the CCPi Regularisation Toolkit~\cite{kazantsev2019ccpi}. TNV was solved using FISTA employing the CIL plugin for the CCPi Regularisation Toolkit~\cite{kazantsev2019ccpi} for the TNV proximal operator. TV-TGV was implemented directly in CIL and solved using PDHG.

Regularisation parameters were carefully chosen to achieve both noise suppression and feature preservation in both spatial and spectral dimensions. For TNV reconstruction, the regularisation parameter $\alpha$ was set to 0.01, while TV-TGV parameters were set to $\beta = 0.0075$ and $\gamma = 0.3$. In both cases, we ran 2000 iterations of the reconstruction algorithm.

\subsection{Post-processing and analysis} \label{subsec:postprocessing}

Quantitative analysis of the reconstructed spectra commonly includes detection and characterisation of the Bragg edges. The Bragg edge positions (in terms of $d$-spacing) allows compositional mapping, as each crystalline structure will have unique set of lattice spacings and hence fingerprint in the neutron transmitted spectrum. The shape of the detected Bragg edges, \emph{i.e.} deviation from abrupt step-like theoretical edge, supports characterisation of crystallographic properties. 

In~\cite{Liptak_2019} authors developed a preprocessing and Bragg edge analysis tool called BEAn for wavelength-resolved neutron transmission data. We adopted Bragg edge detection and fitting functionality from BEAn for the present study. The resulting Bragg edge fitting routine consists of the following steps. (1) BEAn has been developed for transmission data, therefore prior to the analysis we convert reconstructed attenuation data to transmission. (2) The spectrum is then smoothed using the Savitzky-Golay filter, differentiated, smoothed again and passed to a peak-finding routine to detect possible edge locations. Note, smoothing is performed only to support the peak-finding routine and all further steps are performed on unsmoothed spectra. (3) For each detected Bragg edge, a model function derived in~\cite{tremsin2008energy} is fitted using the non-linear least squares fitting method. Since the model function is highly non-linear, the fitting procedure is very sensitive to initialisation parameters. Therefore we use brute force approach to help the fitting procedure to achieve acceptable fit. That is, initialisation parameters are drawn from uniform random distributions chosen to cover a realistic range for each parameter in the model function. The best fit is chosen based on minimum root mean squared error between the fitted function and experimental data.

Additionally, the spatial distribution of individual materials in a sample can be obtained by decomposing the reconstructed spectral images into individual material maps. Material decomposition in spectral CT is an ill-posed problem by its own and various approaches have been proposed in the literature, mainly for spectral X-ray CT. The first group of methods performs material decomposition in the sinogram domain, following by reconstruction of material specific sinograms~\cite{roessl2007k,schlomka2008experimental}. The second group of methods employs simultaneous reconstruction and material decomposition~\cite{barber2016algorithm}. A third approach is to perform material decomposition in the image domain, \emph{i.e.} after the reconstruction step~\cite{firsching2006quantitative,xie2019material}. Limitations and merits of the individual methods are beyond the scope of this research. As a proof of principle, we adapt the later approach here. In particular, we use the so called volume conservation principle~\cite{liu2009quantitative,ronaldson2012toward}, where each voxel in the obtained material maps corresponds to a dimensionless volume fraction occupied by the corresponding material with the full voxel corresponding to a unit volume. This is a valid assumption here as the materials employed in the current study do not mix. Under the volume conservation assumption, the voxel-wise sum of all material maps has to be equal to 1 in each voxel. 

Let $\hat{u}$ be a $K \times D^2$ matrix version of $u$ and let $M$ be the material basis $K \times S$ matrix built up from the predicted wavelength-dependent neutron spectra (\cref{fig:neutron_transmission}), with $S = 6$ being the number of materials employed in this study (5 metals + air). Then, volume fractions of every material $v$, consisting of $S$ row vectors with $D^2$ elements, can be obtained by solving:

\begin{equation} \label{eq:material_decomposition}
	\hat{u} \approx Mv.
\end{equation}

To solve \cref{eq:material_decomposition}, we formulate the following optimisation problem:

\begin{equation} \label{eq:material_decomposition_2}
\begin{aligned}
	\min_{v} \quad & \| \hat{u} - Mv \|^2_F \\
	\textrm{s.t.} \quad & v \geq 0 \\
	& \mathbf{1}_S^T v = \mathbf{1}_{D^2}^T \\
\end{aligned}
\end{equation}

\noindent where $\| \cdot \|^2_F$ is the Frobenius norm and $\mathbf{1}^T$ is a row vector of all ones of appropriate dimension. The first constraint is an element-wise non-negativity constraint enforcing the volume fractions $v$ to be greater than or equal to 0; the second constraint expresses the volume preservation principle. We use CVX~\cite{grant2008cvx,gb08} with the MOSEK solver to solve this optimisation problem.

\section{Results} \label{sec:results}

\subsection{Visual qualitative assessment}

\subsubsection{White beam reconstruction}

\begin{figure}[h!]
	\centering
	\includegraphics[width=4in,height=4in,keepaspectratio]{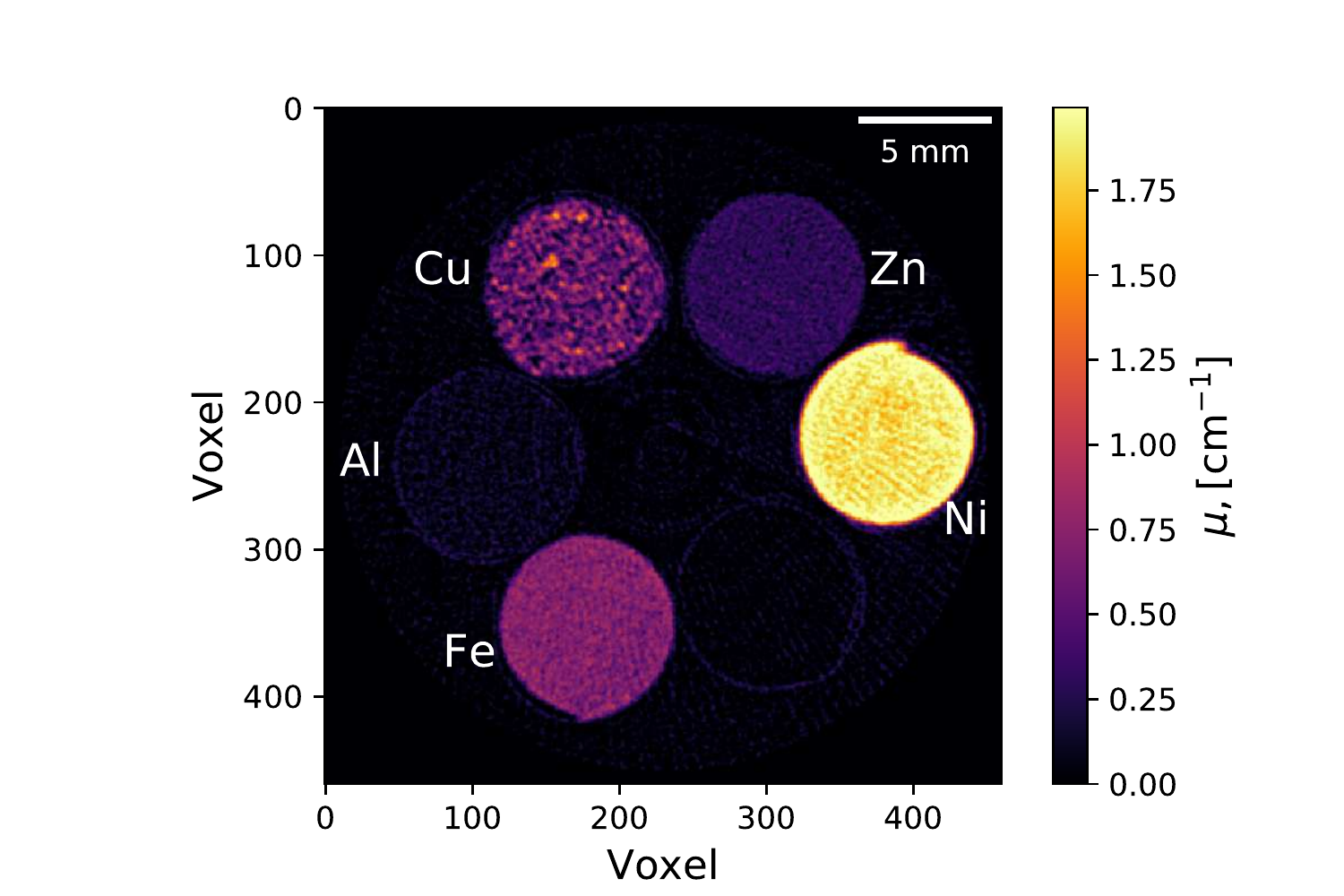}
	\caption{White beam (sum of all wavelength channels) reconstruction using the conventional FBP method.}
	\label{fig:fbp_recon}
\end{figure}

In order to give a visual reference for the spatial dimension, we first perform conventional FBP reconstruction (implemented as an eponymous processor in CIL~\cite{jorgensen2021cil}) of the white beam sinogram (sum of all wavelength channels), as shown in~\cref{fig:fbp_recon}. All cylinders are seen clearly defined and with different contrast for each element. The copper powder used in this study is known to have an average particle size comparable to the voxel size. Therefore reconstruction of the copper cylinder appears inhomogeneous, \emph{i.e.} some reconstructed voxels contain a mixture of copper and air, while some are fully occupied by larger copper particles. There are also noticeable roundness deviations in the reconstructed cross-section of the cylinders. These deviations are caused by the simple fabrication process used for making the phantom. There are also small ``bumps'' where a small portion of powder penetrated overlapping foil layers.

\subsubsection{The spatial dimension}

\begin{figure}[h!]
	\centering
	\includegraphics[clip,trim=0 0in 0.75in 0in,width=\textwidth,keepaspectratio]{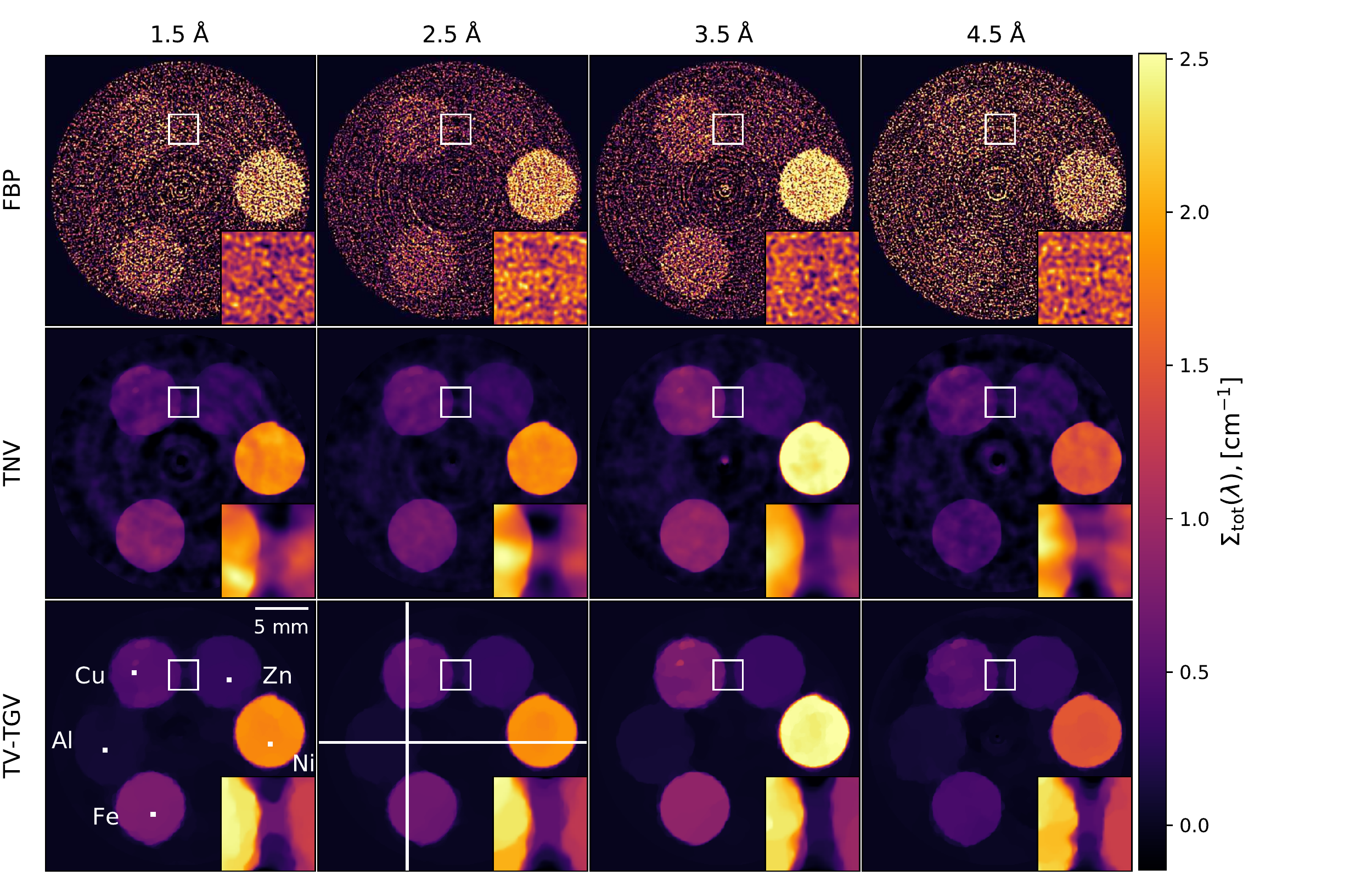}
	\caption{Two dimensional reconstructions of selected (individual) wavelength channels. Top to bottom: channel-wise FBP reconstruction, iterative reconstruction with TNV regularisation and iterative reconstruction with TV-TGV regularisation. A magnified region-of-interest marked with a white rectangle is shown inset. White lines mark profile lines chosen for examination in the next figure, whereas white dots mark individual voxels chosen for spectral comparison in the next section. All slices are visualised using a common colour range. The colour range in the magnified insets is scaled individually to cover only the voxel values in the corresponding inset.}
	\label{fig:recon_comparison}
\end{figure}

\begin{figure}[h!]
	\centering
	\includegraphics[width=\textwidth,keepaspectratio]{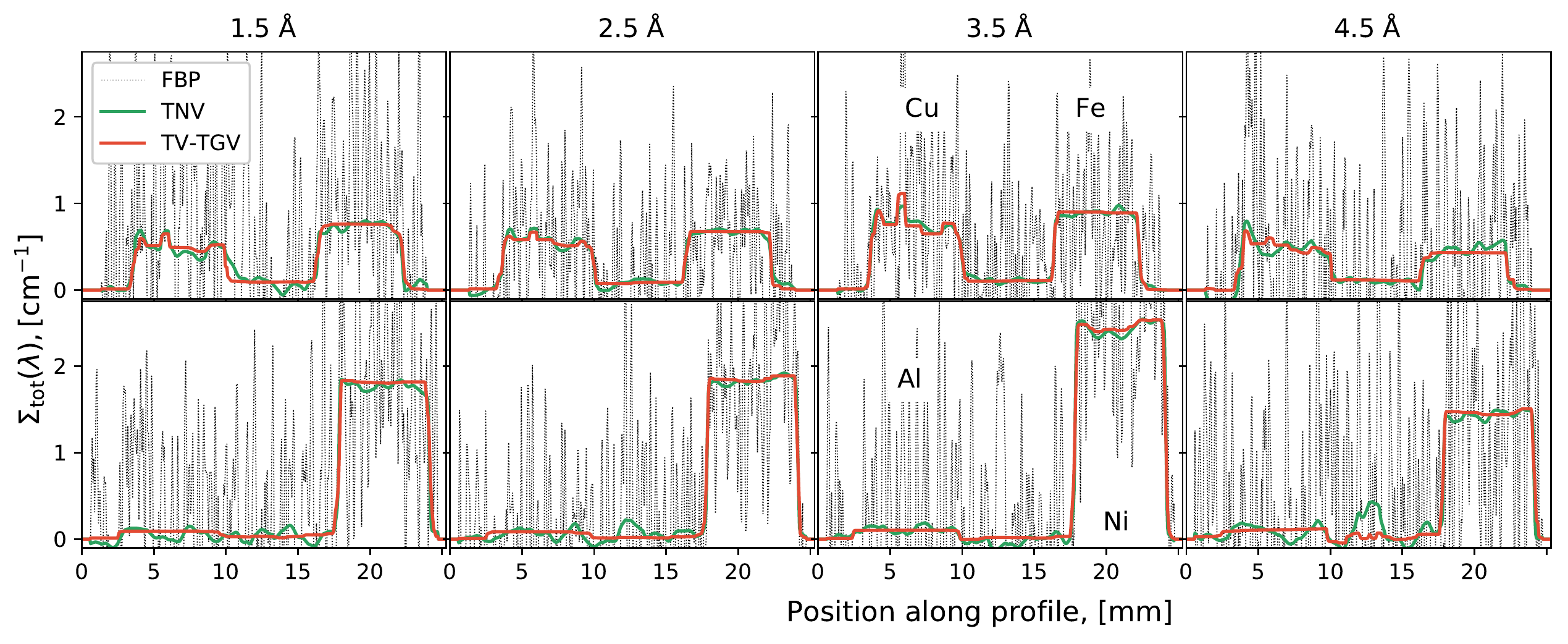}
	\caption{Linescans corresponding to the vertical (top row) and horizontal (bottom row) white lines in~\cref{fig:recon_comparison} (bottom second left) passing through the Cu and Fe cylinders and the Al and Ni cylinders, respectively}
	\label{fig:recon_comparison_profile}
\end{figure}

\Cref{fig:recon_comparison} shows two-dimensional slices for selected (individual) wavelength channels reconstructed using the conventional FBP approach and using the regularised iterative methods discussed in this paper. Linescans across the reconstructed cylinders in \cref{fig:recon_comparison_profile} offer a detailed comparison between the reconstruction methods. We only perform two-dimensional reconstructions in the spatial dimension in this study but results can be generalised to the third dimension. We have chosen a slice roughly in the middle of the upper half of the detector plate in order to avoid detector rows near to the top of the detector or the gap between the detector read-out chips (\cref{subsec:settings}). 

As expected, at these signal levels the FBP reconstruction is barely interpretable (\cref{fig:recon_comparison}, top row). Both TNV and TV-TGV demonstrate drastic improvement in reconstruction quality and noise suppression (\cref{fig:recon_comparison,fig:recon_comparison_profile}). A region-of-interest between two cylinders highlights significant smearing of features in the TNV reconstruction (\cref{fig:recon_comparison}, middle row); the same features appear sharper in the TV-TGV reconstruction (\cref{fig:recon_comparison}, bottom row). TNV suffers from the same limitations as conventional TV-based regularisation: it might oversmooth images. Therefore, TNV suppresses noise and ring artifacts visible in FBP reconstruction but produces blurred and enlarged rings especially prominent in shorter and longer wavelength channels, where counts are much lower. Aluminium has very low neutron attenuation and is invisible in the TNV reconstruction due to contrast loss; another known drawback of both TV and TNV regularisation methods. 

The results demonstrate that the TV-TGV reconstruction is more effective for low contrast features. Furthermore, the TV-TGV reconstruction does not suffer from ring artefacts and the Al cylinder is visible in the reconstructed slices and profile lines (\cref{fig:recon_comparison_profile}, bottom row). Fine features inside the copper cylinder are also partially preserved in the TV-TGV reconstruction (\cref{fig:recon_comparison_profile}, top row).

\subsubsection{The spectral dimension}

\begin{figure}[h!]
	\centering
	\includegraphics[width=\textwidth,keepaspectratio]{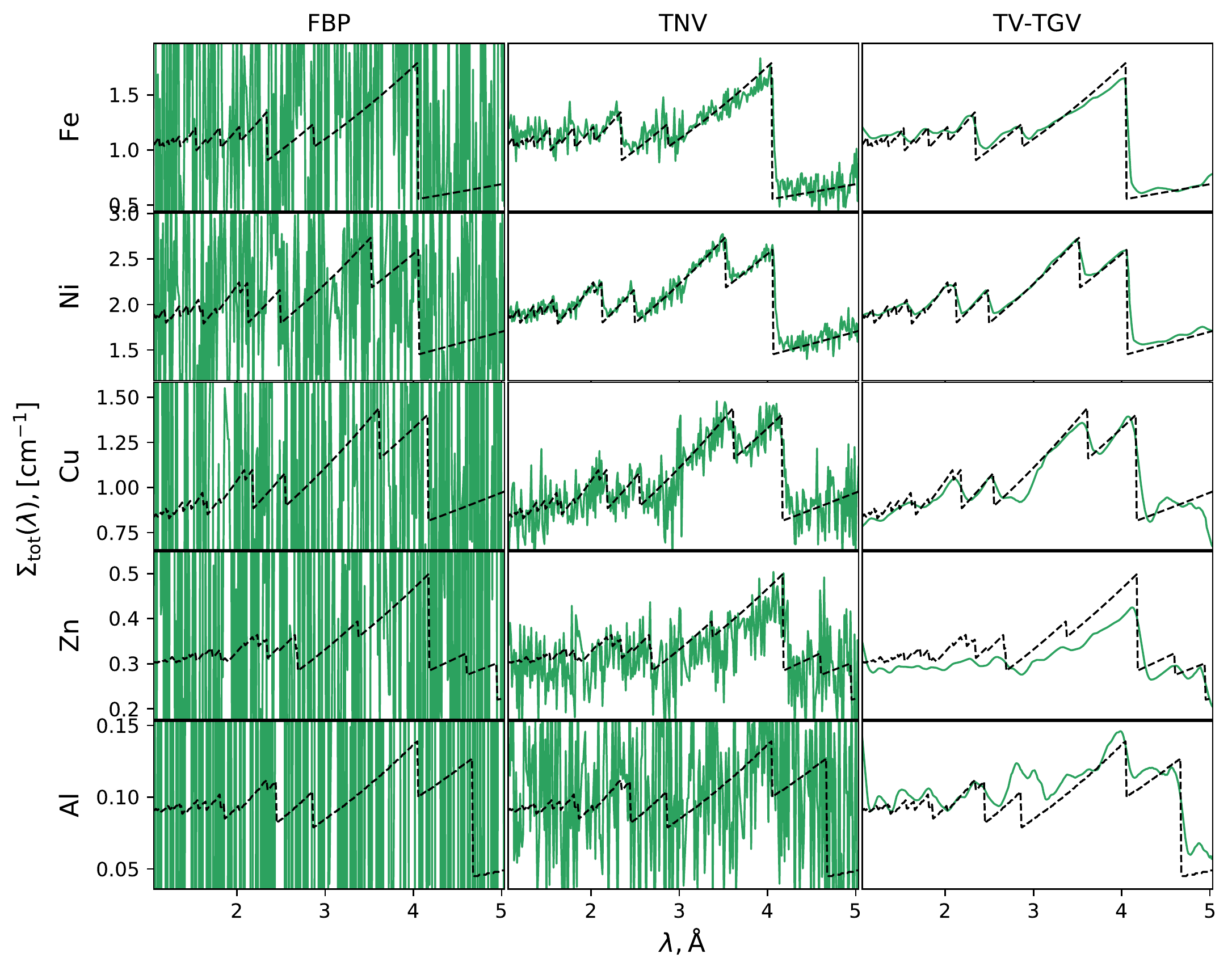}
	\caption{Individual spectra (solid green line) reconstructed by FBP, TNV and TV-TGV for one representative $0.055^3~\mathrm{mm}^3$ voxel located within in each material alongside the predicted signal (dotted black line).}
	\label{fig:spectral_comp}
\end{figure}

The ground truth for the spectral dimension is taken to be the predicted neutron cross-section for the selected materials (\cref{fig:neutron_transmission}). Individual spectra reconstructed for one $0.055^3~\mathrm{mm}^3$ voxel in each of the 5 materials are plotted in~\cref{fig:spectral_comp} alongside the theoretical predictions. The voxel locations inside the cylinders were chosen arbitrarily (marked in~\cref{fig:recon_comparison}).  

The limited counting statistics acquired in our experiment make the FBP reconstructed spectra uninterpretable. By contrast, both the TNV and TV-TGV reconstructions show drastic improvements in reconstruction quality. The reconstructed spectra for Fe, Ni and Cu closely follow the Bragg edge spectra for both TNV and TV-TGV reconstructions. For the low-attenuation materials (Zn and Al), the TV-TGV clearly outperforms TNV. The amplified noise visible in the TNV reconstructions between 2.5~{\AA} and 3~{\AA} is caused by an increase in noise level in the input data, as was highlighted in~\cref{subsec:data_preprocessing}. 

For the high-attenuation materials (Fe, Ni, Cu) both TNV and TV-TGV show comparable performance. While the TV-TGV produces a much smoother spectra, the Bragg edges appear to be less prominent due to smearing (for instance, small edges around 2~{\AA} in Fe). In the case of the TNV regularisation, the noise dominates over smaller Bragg edges. For the low-attenuation materials (Al and Zn), TV-TGV shows better performance as some Bragg edges are visible in the reconstructed spectrum, but are lost in the noise in the case of TNV.

\subsection{Quantitative assessment}

\subsubsection{Mapping of crystallographic information}

\begin{figure}[h!]
	\centering
	\includegraphics[width=\textwidth,keepaspectratio]{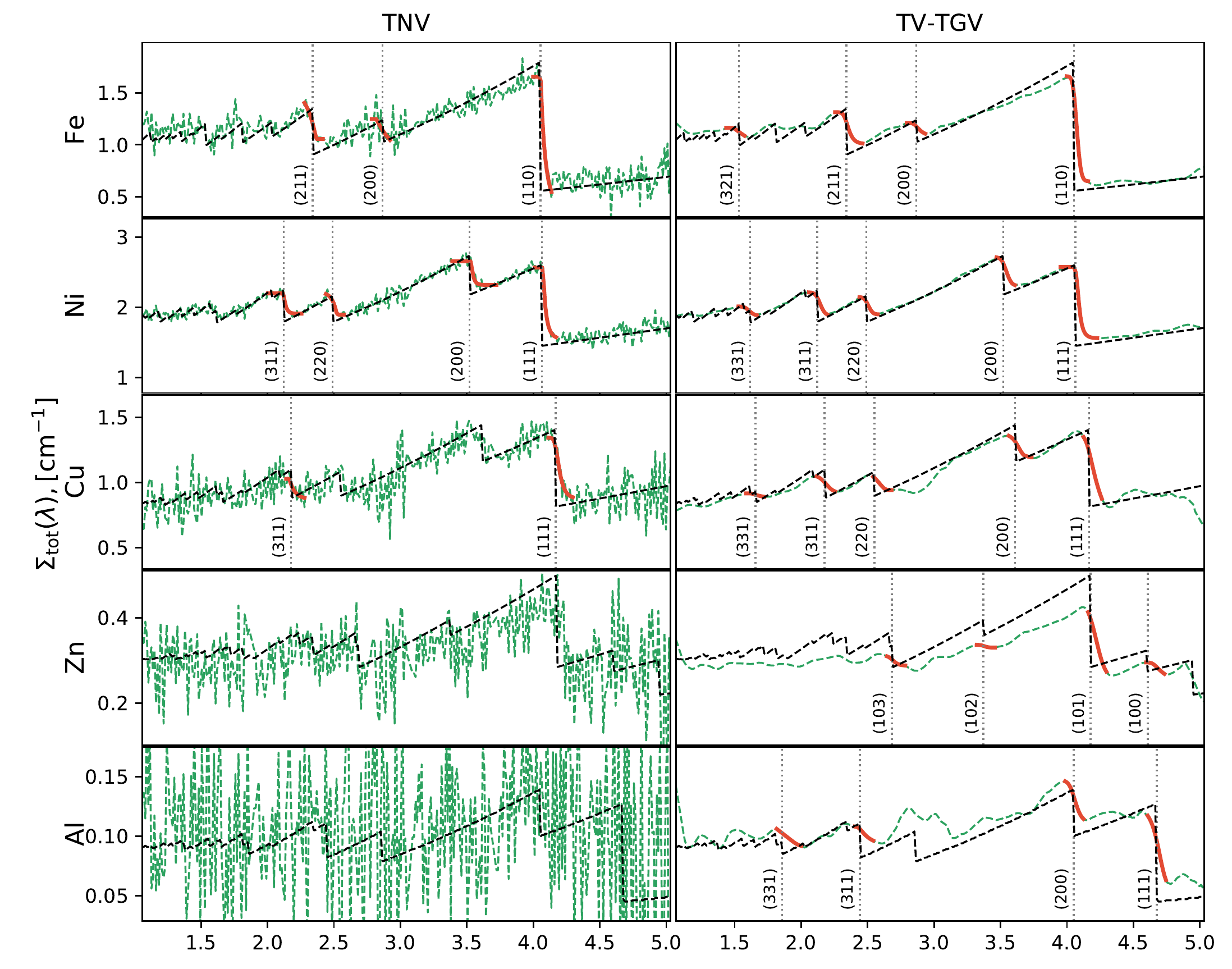}
	\caption{Bragg edge fitting results. Reconstructed spectra (solid green line) are superimposed upon the predicted signal (dotted black line). The model function was fitted around each Bragg edge separately. The fitted model is shown as the solid orange line. Crystalline structural information is included in the reconstructed spectra. }
	\label{fig:bragg_edge_fitting_res}
\end{figure}

Mapping of crystallographic information was performed on individual spectra reconstructed for one $0.055^3~\mathrm{mm}^3$ voxel in each of the 5 materials (the same voxels as in the previous section, marked in~\cref{fig:recon_comparison}). Following the procedure outlined in~\cref{subsec:postprocessing}, the Bragg edges were detected and characterised based on fitting of a dedicated model function (\cref{fig:bragg_edge_fitting_res}). Fitting was performed individually around each detected Bragg edge. The fitted function is shown as the orange solid line. Since the locations of the Bragg edges are directly related to interplanar spacing $d_{hkl}$, we use the reference $\lambda_{hkl} = 2d_{hkl}$ for the corresponding materials to assess the mapping of the crystalline structural information (\cref{table:bragg_edge_fitting}). We also show an absolute error $\varepsilon = \| \lambda_{hkl} - \lambda'_{hkl}\|$ in the estimated $d$-spacing $\lambda'_{hkl}$. For some Bragg edges, the edge detection routine identified edges, but the least squares fitting procedure did not converge to a good solution. We use ``(-)'' to designate such edges. 

\begin{table}[h!]
	\centering
	\begin{tabular}{ c | c | c | c | c | c  }
		\toprule
		& \multirow{2}{*}{$\lambda_{hkl}$} & \multicolumn{2}{ |c| }{TNV} & \multicolumn{2}{ |c }{TV-TGV} \\
		\cline{3-6} 
		& & $\lambda^{\prime}_{hkl}$ & $\varepsilon$& $\lambda^{\prime}_{hkl}$ & $\varepsilon$ \\
		\midrule
		\midrule
		Fe(110) & 4.054 & 4.055 & 0.001 & 4.077 & 0.023 \\
		Fe(200) & 2.866 & 2.819 & 0.047 & 2.853 & 0.013 \\
		Fe(211) & 2.340 & 2.366 & 0.026 & 2.332 & 0.008 \\
		Fe(220) & 2.026 & \multicolumn{2}{ |c| }{-} & \multicolumn{2}{ |c }{(-)}\\
		Fe(310) & 1.812 & \multicolumn{2}{ |c| }{(-)} & \multicolumn{2}{ |c }{-} \\
		Fe(222) & 1.654 & \multicolumn{2}{ |c| }{-} & \multicolumn{2}{ |c }{-} \\
		Fe(321) & 1.532 & \multicolumn{2}{ |c| }{-} & 1.492 & 0.04 \\
		\midrule
		Ni(111) & 4.064 & 4.073 & 0.009 & 4.076 & 0.012 \\
		Ni(200) & 3.520 & 3.533 & 0.013 & 3.545 & 0.025 \\
		Ni(220) & 2.490 & 2.522 & 0.032 & 2.512 & 0.022 \\
		Ni(311) & 2.122 & 2.122 & 0.000 & 2.136 & 0.014 \\
		Ni(222) & 2.032 & \multicolumn{2}{ |c| }{-} & \multicolumn{2}{ |c}{-} \\
		Ni(400) & 1.760 & \multicolumn{2}{ |c| }{-} & \multicolumn{2}{ |c }{-} \\
		Ni(331) & 1.616 & \multicolumn{2}{ |c| }{(-)} & 1.609 & 0.007 \\
		\midrule
		Cu(111) & 4.168 & 4.216 & 0.048 & 4.151 & 0.017 \\
		Cu(200) & 3.610 & \multicolumn{2}{ |c| }{(-)} & 3.662 & 0.052 \\
		Cu(220) & 2.552 & \multicolumn{2}{ |c| }{(-)} & 2.627 & 0.075 \\
		Cu(311) & 2.176 & 2.163 & 0.013 & 2.215 & 0.039 \\
		Cu(222) & 2.084 & \multicolumn{2}{ |c| }{-} & \multicolumn{2}{ |c }{-} \\
		Cu(400) & 1.804 & \multicolumn{2}{ |c| }{-} & \multicolumn{2}{ |c }{-} \\
		Cu(331) & 1.656 & \multicolumn{2}{ |c| }{-}  & 1.629 & 0.027 \\
		\midrule
		Al(111) & 4.676 & \multicolumn{2}{ |c| }{-} & 4.729 &0.053 \\
		Al(200) & 4.050 & \multicolumn{2}{ |c| }{-} & 4.040 &0.01 \\
		Al(220) & 2.864 & \multicolumn{2}{ |c| }{-} & \multicolumn{2}{ |c }{(-)} \\
		Al(311) & 2.442 & \multicolumn{2}{ |c| }{-} & 2.435 & 0.007 \\
		Al(222) & 2.338 & \multicolumn{2}{ |c| }{-} & \multicolumn{2}{ |c }{-} \\
		Al(400) & 2.024 & \multicolumn{2}{ |c| }{-} & \multicolumn{2}{ |c }{-} \\
		Al(331) & 1.858 & \multicolumn{2}{ |c| }{-} & 1.982 & 0.124 \\
		\midrule
		Zn(002) & 4.950 & \multicolumn{2}{ |c| }{-} & \multicolumn{2}{ |c }{-} \\
		Zn(100) & 4.608 & \multicolumn{2}{ |c| }{(-)} & 4.648 & 0.040 \\
		Zn(101) & 4.178 & \multicolumn{2}{ |c| }{(-)} & 4.209 & 0.031 \\
		Zn(102) & 3.372 & \multicolumn{2}{ |c| }{(-)} & 3.391 & 0.019 \\
		Zn(103) & 2.682 & \multicolumn{2}{ |c| }{(-)} & 2.671 & 0.011 \\
		\bottomrule
	\end{tabular} 
	\caption{Results of mapping of crystallographic information. Estimated $\lambda'_{hkl}$ is compared vs. reference $\lambda'_{hkl} = 2 d_{hkl}$, where $d_{hkl}$ were extracted from NXS code~\cite{Boin:db5100}. For some Bragg edges, the edge detection routine identified edges but the least squares fitting procedure did not converge to a good solution. We use ``(-)'' to designate such edges. Edges that could not be detected by the automated procedure are marked with ``-''.}
	\label{table:bragg_edge_fitting}
\end{table}

For the high-attenuation materials in this study, \emph{i.e.} Fe and Ni, both TNV and TV-TGV reconstructions show comparable results in terms of error and uncertainty in estimated interplanar spacing. The lower the material's attenuation coefficient, the more the TV-TGV reconstruction surpasses the TNV results. Thus, in the TNV reconstruction we can estimate only some interplanar distances; no Bragg edges can be detected and characterised for Al. Remarkably, the TV-TGV reconstruction is able to locate some Bragg edges even for the low-attenuation materials such as Zn and Al. Accuracy in estimated interplanar distances deteriorate with prominence of corresponding Bragg edges. Also, Bragg edge fitting performs better for isolated edges because of edges smearing.

Of course one could increase the sensitivity to the low scattering materials by increasing the acquisition period or the number of projections, but only at the expense of longer total experiment time, which may be prohibitive. In this study we also used an automated procedure to detect and fit the model function (\cref{subsec:postprocessing}). A more careful manual procedure might improve the Bragg edge fitting results, especially for the noisier TNV reconstruction.

\subsubsection{Material maps}

\begin{figure}[h!]
	\centering
	\includegraphics[clip,trim=0in 0in 1.25in 0in,width=\textwidth,keepaspectratio]{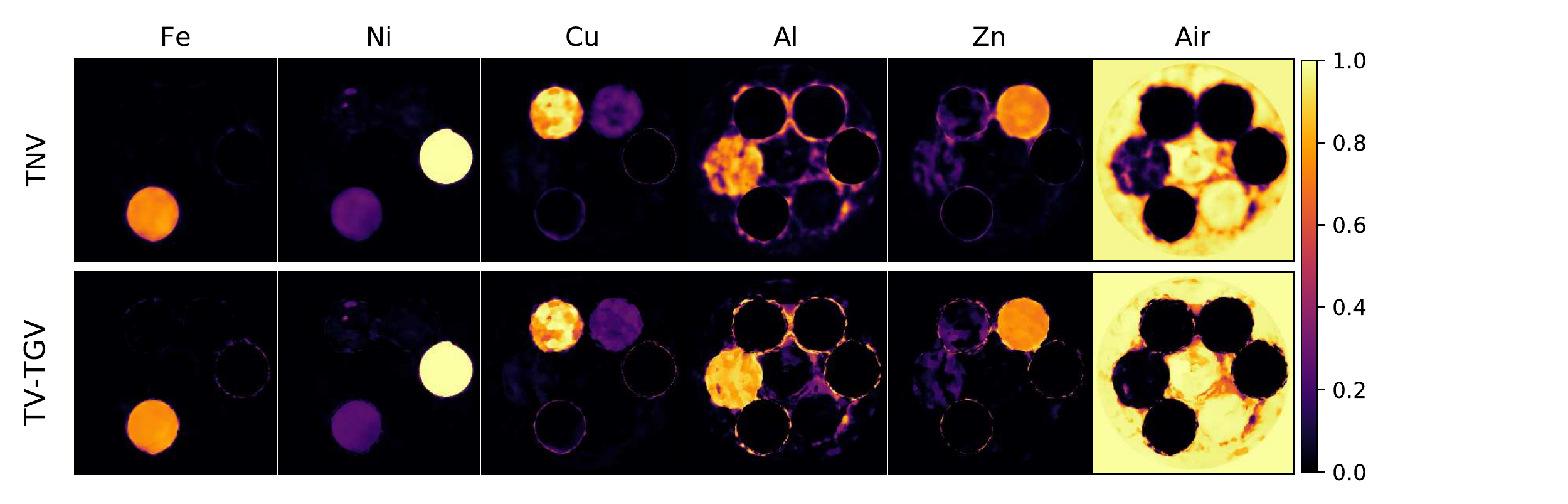}
	\caption{Material maps. Intensity value in each voxel is equal to the voxel volume fraction occupied by the corresponding material (calculated based on the material decomposition method described in~\cref{subsec:postprocessing}).}
	\label{fig:material_map}
\end{figure}

Finally, in~\cref{fig:material_map} we show spatial material maps calculated on the basis of the material decomposition method described in~\cref{subsec:postprocessing}. The difference between the material maps calculated from TNV and TV-TGV reconstructed images is quite subtle, especially for high-attenuation materials. Advantages of the TV-TGV approach are most pronounced in the Al material map. Compared with TNV, TV-TGV does a better job in preserving the fine Al foil containers and the Al cylinder appears more uniform. The material maps for both reconstructions exhibit some artifacts, for instance, the faint Fe cylinder is visible in the Ni map and similarly Zn is visible in the Cu map. The artifacts are caused by the fact that the most prominent Bragg edges almost coincides in the Fe and Ni spectra, as well as in the Cu and Zn spectra (\cref{fig:neutron_transmission}). In this study we used a fairly generic method for material decomposition. As we have already mentioned in~\cref{subsec:postprocessing}, the material decomposition problem is an ill-posed problem and various regularisation methods can be used to improve material decomposition depending on underlying image properties in the spatial and spectral dimensions.

\section{Discussion and conclusions} \label{sec:conclusion}

The unique features of the pulsed time of flight neutron source allow quantification of material properties in bulk samples. Low flux and slow acquisition result in long scan times being required to record sufficient counts across hundreds of wavelength channels, making material characterisation at the full attainable resolution infeasible in practice. Joint reconstruction which exploits inter-channel correlations in multi-channel images is an effective way to shift feature recovery from exposure to reconstruction. In this work we have studied the performance of advanced reconstruction methods with the dedicated multi-channel regularisation techniques for Bragg edge neutron CT. We have demonstrated that the tailored TV-TGV regularisation technique, which favours specific image properties in the spatial and spectral dimensions, allows retrieval of crystallographic properties at a resolution previously unattainable through conventional reconstruction methods using the same exposure time. The proposed technique was compared with the TNV method - a recent regularisation technique developed for spectral CT. While both methods clearly outperform FBP, TV-TGV yields better reconstruction of low-contrast features. Furthermore, quantitative comparisons were also used to evaluate the performance of both methods. We extracted crystallographic properties from the reconstructed spectra based on detection and characterisation of Bragg edges. TV-TGV facilitated extraction of interplanar spacings for all materials employed in this study; no Bragg edges could be characterised for the low-attenuation materials in the TNV reconstruction.

Our study is closely related to, and builds on, the proof-of-concept study presented in~\cite{Carminati:ei5051} using the IMAT beamline, a similar physical phantom (containing Ni, Fe, Cu, Ti, Pb, Al) and channel-wise FBP reconstruction. The authors demonstrated Bragg edge fitting with an absolute error in Bragg edge location within 0.04~{\AA} for Ni ((111), (200), (220), (311)) and Fe ((110), (200), (211)). In the current study however the dataset was acquired with three times shorter exposure time, two times higher spectral resolution and 25 times smaller scattering volume ($0.275 \times 0.275 \times 0.055~\mathrm{mm}^3$ region-of-interest vs. $0.055^3~\mathrm{mm}^3$ voxel) for Bragg edge characterisation. This represents counting statistics some 150 times worse than in the proof-of-concept study yet comparable image quality and edge detection is observed. This is what is achievable by replacing the conventional FBP reconstruction methods with TNV or our proposed TV-TGV method. For the Bragg edges characterised in~\cite{Carminati:ei5051}, the absolute error in Bragg edge location in the present study is within 0.03~{\AA}; for other less prominent Bragg edges, which were not characterised in~\cite{Carminati:ei5051}, error is within 0.1~{\AA}. 

There are two well-known limitations of iterative reconstruction techniques. First, they present a significant computational burden. Depending on implementation and available hardware, reconstruction of a four-dimensional dataset can take a few days. However for ToF neutron CT, where typical exposure time per projection is between 30 minutes and an hour, and where the beamtime cost for every measurement hour is extremely high, slow reconstruction is an acceptable price to pay for increased sample throughput. Secondly, choice of a regularisation parameter, which balances the fitting and regulatisation terms in the reconstruction procedure, is a topic of extensive research in the inverse problems community. Here, we manually tuned regularisation parameters based on visual inspection. Automated procedures to define the regularisations parameters are needed to run the proposed methods on an everyday basis.

The method proposed in this paper is by no means limited to ToF neutron CT but is applicable to other spectral CT modalities. We expect that further quality improvements can be achieved if the full four-dimensional spatio-spectral volume is reconstructed as regularisation along the axial direction will be used to penalise image variations and suppress noise. Furthermore, we plan to join reconstruction and material decomposition, such that the spatial volume fraction maps of basis materials are reconstructed directly, instead of reconstruction of hundreds of channels followed by the material decomposition in the image domain. This approach allows a significant reduction of required computations and storage and also have the potential to further improve image quality through the additional regularisation.

There is an ongoing work on providing the reconstruction methods described in this paper to IMAT scientists and users through MANTID Imaging~\cite{dimitar_tasev_2021_4451979} -- a graphical toolkit for processing neutron imaging and tomography data. Relevant CIL modules~\cite{jorgensen2021cil,papoutsellis2021cil} will be integrated as MANTID Imaging plugins, supporting not only high-resolution ToF neutron CT investigations and a more efficient usage of valuable neutron beamtime but also bridging a gap between theory and applications of advanced reconstruction methods.

\section*{Funding}

\sloppy
This work was funded by EPSRC grants ``A Reconstruction Toolkit for Multichannel CT'' (EP/P02226X/1), ``CCPi: Collaborative Computational Project in Tomographic Imaging'' (EP/M022498/1 and EP/T026677/1). We gratefully acknowledge beamtime RB1820541 at the IMAT Beamline of the ISIS Neutron and Muon Source, Harwell, UK. EA was partially funded by the Federal Ministry of Education and Research (BMBF) and the Baden-W{\"u}rttemberg Ministry of Science as part of the Excellence Strategy of the German Federal and State Governments. JSJ was partially supported by The Villum Foundation (grant no. 25893). WRBL acknowledges support from a Royal Society Wolfson Research Merit Award. PJW acknowledges support from the European Research Council grant No. 695638 CORREL-CT. 

\section*{Acknowledgments}

Authors would like to thank Dr. Winfried Kockelmann (ISIS STFC), Dr. Anton Tremsin (UC Berkeley), Dr. Joe Kelleher (ISIS STFC), Dr. S{\o}ren Schmidt (ESS) and Alexander Liptak (University of Liverpool) for the fruitful discussions on challenges in neutron CT and various approaches to overcome them. This work made use of computational support by CoSeC, the Computational Science Centre for Research Communities, through the Collaborative Computational Project in Tomographic Imaging (CCPi). 

\bibliographystyle{vancouver}
\bibliography{neutron}
 
\end{document}